\documentclass[aps,10pt,letterpaper,twocolumn,english,showpacs,preprintnumbers,amsmath,amssymb,nofootinbib,prl]{revtex4-1}
\usepackage{amsmath}
\usepackage{graphicx}
\usepackage{setspace}
\usepackage{amssymb}
\usepackage{dcolumn}
\usepackage{mathrsfs}
\usepackage{bm}
\usepackage{longtable}

\begin{document}

\preprint{}

\title{Narrowband Biphotons with Polarization-Frequency Coupled Hyperentanglement}

\author{Chi Shu}
\author{Xianxin Guo}
\author{Peng Chen}
\author{M. M. T. Loy}
\author{Shengwang Du} \email{dusw@ust.hk}
\affiliation{Department of Physics, The Hong Kong University of Science and Technology, Clear Water Bay, Kowloon, Hong Kong, China}

\date{\today}

\begin{abstract}
We demonstrate the generation of narrowband biphotons with polarization-frequency coupled hyperentanglement from spontaneous four-wave mixing in cold atoms. The coupling between  polarization and frequency is realized through a frequency shifter and linear optics. When the polarization-frequency degrees of freedom are decoupled, it is robust to create polarization and frequency Bell states, confirmed by the polarization quantum-state tomography and the two-photon temporal quantum beating. Making use of the polarization-frequency coupling to transfer polarization phase retard to the entangled frequency modes, we produce a frequency Bell state with tunable phase difference between its two bases.
\end{abstract}

\pacs{03.67.Bg, 42.50.Dv, 03.65.Aa, }


\maketitle


Entangled photon pairs, termed \textit{biphotons}, have become benchmark tools not only for probing fundamental quantum physics but also for realizing applications in quantum communication and quantum computation \cite{PanRMP2012}. Today, photonic entanglement can be generated in many degrees of freedom, including polarization \cite{PolarizationE1988Ou, PolarizationE1988Shih, PolarizationE1995}, path (momentum) \cite{PathE1989, PathE1990}, orbital angular momentum \cite{OAME2001}, and time (frequency)\cite{FransonPRL1989,BrendelPRL1999}. Although entanglement is a general property of a multipartite quantum system, Bell-type entangled states are particularly important for quantum information processing and quantum teleportation \cite{Nielsen2000}. For wide-band biphotons generated from spontaneous parametric down conversion, Bell states of temporal entanglement can be obtained using Franson interferometry \cite{FransonPRL1989,BrendelPRL1999}, and frequency-entangled qubits can be realized by shaping the energy spectrum \cite{BernhardPRA2013} and polarization entanglement transfer \cite{ZeilingerPRL2009}. So far, frequency Bell states for narrowband (1-50 MHz) biphotons have only been observed from spontaneous four-wave mixing (SFWM) in cold atoms \cite{DuJOSAB2008, DuPRL2011}. However, in such a system, it is difficult to manipulate the frequency entanglement that is naturally endowed by energy conservation.

In this Letter, we demonstrate a robust scheme for generating narrowband biphotons with polarization-frequency coupled hyperentanglement. Continuous-wave SFWM in laser cooled atoms has been demonstrated as an efficient method of producing biphotons with bandwidth comparable to atomic natural linewidth ($\sim$10 MHz) \cite{HarrisPRl2005, VuleticScience2006, Subnatural, KurtsierferPRL2013, DuOptica2014} which are of great interest for realizing efficient photon-atom quantum interfaces in a quantum network \cite{QuantumInternet}. However, existing methods for generating polarization entanglement in such a system are complicated and requires special technical cares \cite{DuPRL2011, YanPRL2014}. In our simple scheme, after the generation of biphotons from SFWM, we create hyperentanglement in both polarization and frequency simultaneously making use of an acousto-optical modulator (AOM) and linear optics. When the hyperentanglement is decoupled, we obtain controllable Bell states in either polarization or frequency modes.

\begin{figure}
\includegraphics[width=\linewidth]{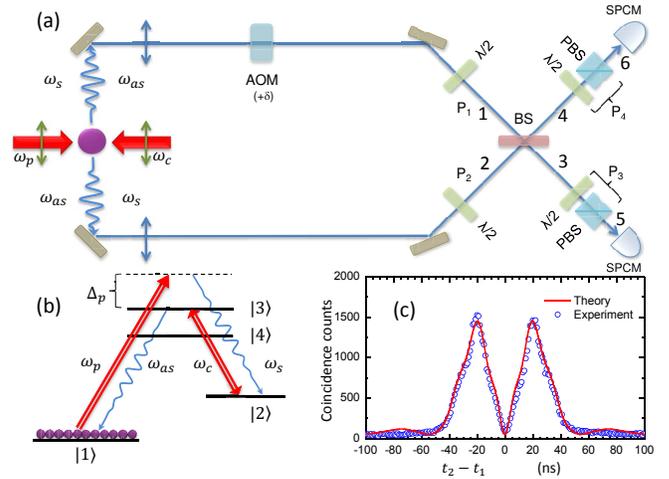}
\caption{\label{fig:ExperimetalSetup} (color online). (a) Experimental schematic of narrowband biphotons generation and detection with polarization-frequency coupled hyperentanglement. The photon pairs are produced from SFWM in cold $^{85}$Rb atoms. (b) $^{85}$Rb atomic energy level diagram. (c) Two-photon coincidence counts before the BS as a function of relative time delay between the ports 1 and 2.}
\end{figure}

Our experimental configuration is illustrated in Fig.~\ref{fig:ExperimetalSetup}(a). We produce narrowband photon pairs from SFWM in laser-cooled $^{85}$Rb atoms driven by two coherent (pump and coupling) laser fields \cite{DuPRA2012}. The relevant atomic energy levels in Fig.~\ref{fig:ExperimetalSetup}(b) are $|1\rangle=|5S_{1/2}, F=2\rangle$, $|2\rangle=|5S_{1/2}, F=3\rangle$, $|3\rangle=|5P_{3/2}, F=3\rangle$, and $|4\rangle=|5P_{3/2}, F=2\rangle$. The coupling laser ($\omega_c$, 3 mW, diameter 2mm) is resonant to transition $|2\rangle\leftrightarrow|3\rangle$. The pump laser ($\omega_p$, 40 mW, diameter 2 mm) is far blue detuned from transition $|1\rangle\rightarrow|3\rangle$ so that the majority of the atomic population remains in the ground level $|1\rangle$. In presence of counter-propagating pump and coupling laser beams, phase-matched backward paired Stokes ($\omega_s$) and anti-Stokes ($\omega_{as}$) photons are produced into two symmetric paths (1 and 2) at right angle with respect to the pump-coupling beams: Stokes photons go to port 1 and anti-Stokes photons go to port 2, and vice versa. The atomic optical depth on the anti-Stokes transition is 5. The photons in path 1 pass through an AOM (Brimrose) and their angular frequencies are up-shifted by $\delta=2\pi\times100$ MHz. We collect the photons from the atomic source at at the horizontal polarization with two linear polarizers. Their polarizations are later rotated to P$_1$ and P$_2$ by two half-wave ($\lambda/2$) plates. To measure the two-photon coincidence counts, the photons are coupled into two single-mode fibers that are connected to two single-photon counting modules (SPCM, Excelitas SPCM-AQRH-16-FC). Figure \ref{fig:ExperimetalSetup}(c) shows the two-photon correlation between paths 1 and 2. The coincidence counts are collected over 3900 s and analyzed by a time-to-digital converter (Fast Comtec P7888) with a time bin width of 1 ns. Since in each photon pair the Stokes photon is always produced before its paired anti-Stokes photon and they never appear at the same time \cite{DuPRA2007}, we can distinguish the Stokes and anti-Stokes photons with time-resolved correlation measurement without frequency filtering. The signal at $t_2-t_1>0$ represents the Stokes (path 1) to anti-Stokes (path 2) correlation: $G^{(2)}_{0}(t_2-t_1)=|\langle t_1, t_2|\omega_s+\delta,\omega_{as}\rangle|^2=|\psi_0(t_2-t_1)|^2$. The signal at $t_2-t_1<0$ represents the anti-Stokes (path 1) to Stokes (path 2) correlation: $G^{(2)}_{0}(t_1-t_2)=|\langle t_1, t_2|\omega_{as}+\delta,\omega_{s}\rangle|^2=|\psi_0(t_1-t_2)|^2$. Here $\psi_0(\tau)$ is the Stokes-anti-Stokes biphotons relative waveform from the SFWM source. The experimental data agrees well with the theoretical curve obtained numerically following the interaction picture \cite{DuJOSAB2008}. The biphotons have correlation time about 50 ns. We also perform a temporal quantum-state tomography \cite{TDQST} and obtain the biphotons bandwidth of about 22 MHz.

The paths 1 and 2 are then combined by a 50:50 beam splitter (BS). In our setup, there is no relative length difference between the paths 1 and 2. The two-photon state after the BS with Stokes photon in the output port 3 and anti-Stokes photon in the output port 4 is described as
\begin{eqnarray}
|\Psi_{s3,as4}\rangle=\frac{1}{\sqrt{2}}\big(|P_1, \omega_{s}+\delta\rangle_3|P_2,\omega_{as}\rangle_4 \nonumber \\
-|P_2, \omega_{s}\rangle_3|P_1, \omega_{as}+\delta\rangle_4\big). \label{eq:Hyperentanglement0}
\end{eqnarray}
Setting P$_1$=H (horizontal polarization) and P$_2$=V (vertical polarization), we obtain the following hyperentangled state
\begin{eqnarray}
|\Psi_{s3,as4}\rangle=\frac{1}{\sqrt{2}}\big(|H,\omega_{s}+\delta,\rangle_3|V,\omega_{as}\rangle_4 \nonumber \\
-|V,\omega_{s}\rangle_3|H,\omega_{as}+\delta\rangle_4\big), \label{eq:Hyperentanglement}
\end{eqnarray}
where the frequency modes and polarization states are coupled together and inseparable.

\begin{figure}
\includegraphics[width=\linewidth]{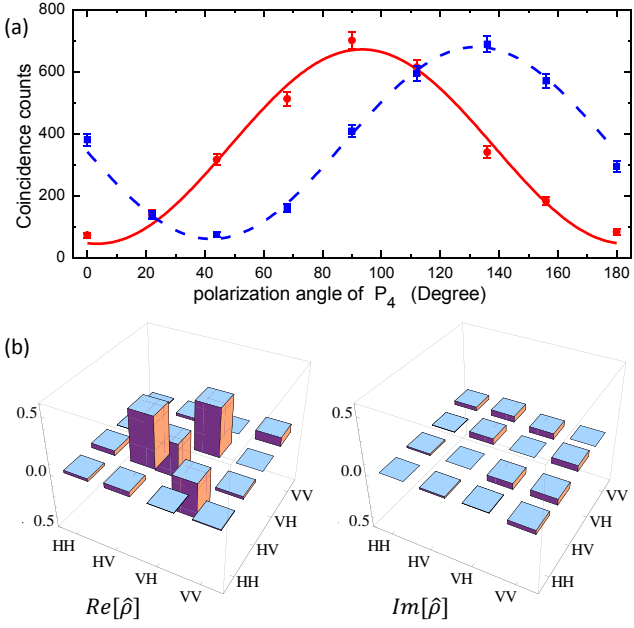}
\caption{\label{fig:PolarizationEntanglement} (color online). Polarization entanglement under the decoupled condition $\delta=0$. (a) Polarization correlation between the paired photons. The coincidence counts are integrated over a 90 ns coincidence window that covers the entire biphotons wave packet. The polarization angle of the linear polarizer P$_3$ are fixed at 0 and 45$^o$ to the horizontal axis for the circular and square experimental data, respectively. (b) Real and imaginary parts of the polarization state density matrix reconstructed from the polarization quantum-state tomography.}
\end{figure}

To show the robustness of the scheme in entanglement manipulation, we first decouple the frequency-polarization degrees of freedom by removing the AOM from path 1, \textit{i.e.} $\delta=0$. Then the state in Eq.~(\ref{eq:Hyperentanglement}) is reduced to
\begin{eqnarray}
\frac{1}{\sqrt{2}}\big(|HV\rangle-|VH\rangle\otimes|\omega_s,\omega_{as}\rangle,
\label{eq:PE}
\end{eqnarray}
which is one of the polarization-entangled Bell states. To verify the polarization entanglement, we add linear polarizers P$_3$ and P$_4$ at the two BS output ports 3 and 4, respectively. Each polarizer consists of a half-wave plate and a cubic polarizing beam splitter (PBS), as shown in Fig.~\ref{fig:ExperimetalSetup}(a). The measured two-photon polarization correlations are displayed in Fig.~\ref{fig:PolarizationEntanglement}(a). We take the circular (square) data points by fixing P$_3=H$ [or $(H+V)/\sqrt2$] and varying the angle of P$_4$. The solid sinusoidal curves are the best fits with visibilities of $87\pm7$\% and $84\pm5$\% respectively, which violate the Bell-CHSH inequality and confirm the state entanglement \cite{CHSH}. we further perform a polarization quantum-state tomography to fully characterize the obtained state \cite{PQST_1,PQST_2}. To do this, we insert a quarter-wave plate before each polarizer (P$_3$, P$_4$). The density matrix is reconstructed from 16 independent coincidence counting measurements using the maximum likelihood estimation method:
\[\left({\begin{array}{cccc}
\scriptstyle 0.02&\scriptstyle\ 0.04+0.01i&\scriptstyle\ 0.00&\scriptstyle 0.01+0.03i\\
\scriptstyle 0.04-0.01i&\scriptstyle\ 0.46&\scriptstyle\ -0.33+0.05i&\scriptstyle 0.02+0.05i\\
\scriptstyle 0.00&\scriptstyle\ -0.33-0.05i&\scriptstyle\ 0.44&\scriptstyle -0.05i\\
\scriptstyle 0.01-0.03i&\scriptstyle\ 0.02-0.05i&\scriptstyle\ 0.05i&\scriptstyle 0.08\\
\end{array}}\right)\]
whose graphical representation is shown in Figure \ref{fig:PolarizationEntanglement}(b). The fidelity between the measurement and the ideal state in Eq.~(\ref{eq:PE}) is 88.3\%. We use the obtained density matrix to test the violation of the Bell-CHSH inequality ($|S|<2$) and obtain $S=2.2\pm0.1$. Once we obtain this state, we can produce the other three independent Bell states using additional birefringent phase shifters (such as wave plates) \cite{ZeilingerPRL1996}. The method demonstrated here is much simpler than that in the recent work \cite{YanPRL2014}, and it does not require any phase stabilization.
\begin{figure}
\includegraphics[width=\linewidth]{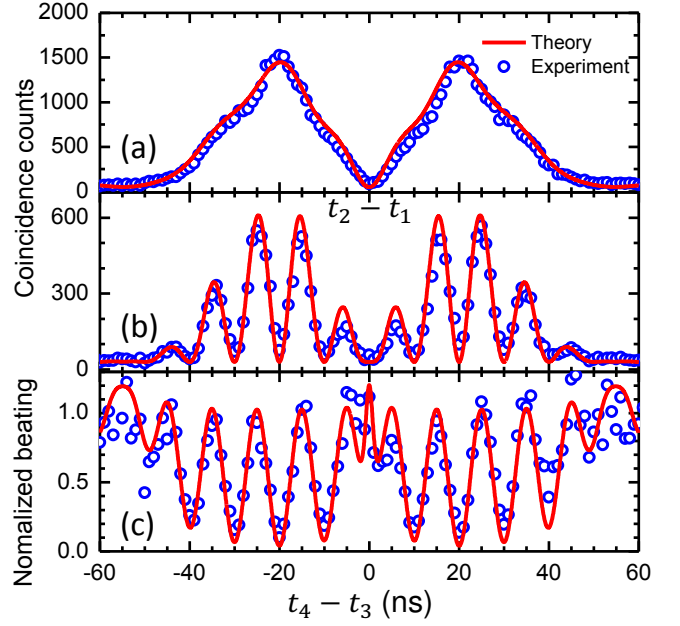}
\caption{\label{fig:FEBeating} (color online). Frequency entanglement under the decoupled condition P$_1$=P$_2$=P$_0$. (a) Biphotons waveform before the BS. (b) Two-photon interference coincidence counts measured at the output ports 3 and 4 of the BS. (c) Normalized two-photon beating signal.}
\end{figure}

We can also decouple the frequency-polarization degrees of freedom by setting P$_1$=P$_2$=P$_0$ that reduces Eq.~(\ref{eq:Hyperentanglement}) to a frequency-entangled Bell state
\begin{eqnarray}
|P_0 P_0\rangle\otimes\frac{1}{\sqrt{2}}\big(|\omega_{s}+\delta,\omega_{as}\rangle-|\omega_{s},\omega_{as}+\delta\rangle\big).
\label{eq:FE}
\end{eqnarray}
The two-photon wave function in time domain can be derived as
\begin{eqnarray}
\Psi(t_3,t_4)=ie^{-i(\omega_{s}+\delta/2)t_3}e^{-i(\omega_{as}+\delta/2)t_4}\psi_0(\tau)\sin(\delta \tau/2),\nonumber\\
\label{eq:FEwave}
\end{eqnarray}
where $\tau=t_4-t_3$. We then obtain the Glauber correlation function
\begin{eqnarray}
G^{(2)}_{34}(\tau)=|\Psi(t_3,t_4)|^2=\frac{1}{2}G^{(2)}_0(\tau)\big[1-\cos(\delta \tau)\big],
\label{eq:FEG2}
\end{eqnarray}
where $G^{(2)}_0(\tau)$ is the Stokes-anti-Stokes two-photon Glauber correlation function before the BS. Equation (\ref{eq:FEG2}) displays a two-photon quantum beating with the frequency of $\delta/(2\pi)$ associated with the envelop of $G^{(2)}_0(\tau)$. With the two-photon joint detection efficiency $\eta$, the time-bin width $\Delta t$, and the total measurement time $T$, the two-photon coincidence counts can be calculated as $C_{34}(\tau)=G_{34}^{(2)}(\tau)\eta\Delta t T$. The experimental result of quantum beating is shown in Fig.~\ref{fig:FEBeating}. For comparison, we plot the two-photon coincidence counts without interference in Fig.~\ref{fig:FEBeating}(a), measured before the BS, the same as Fig.~\ref{fig:ExperimetalSetup}(c). Figure \ref{fig:FEBeating}(b) displays the two-photon beating, as predicted from Eq.~(\ref{eq:FEG2}). To see the beating more clearly, we normalize the two-photon interference to its envelop in Fig.~\ref{fig:FEBeating}(a) and plot the normalized two-photon beating in Fig.~\ref{fig:FEBeating}(c). By besting fitting the data with a sinusoidal wave we determine the visibility V=$80\pm2$\%, which is far beyond the requirement for violation of Bell inequality in time-frequency domain\cite{FransonPRL1989}.

Starting from the state in Eq.~(\ref{eq:Hyperentanglement}) and following the same procedures as Ref. \cite{ZeilingerPRL1996}, one can obtain the following 8 independent polarization-frequency coupled, hyperentangled quantum states by placing additional wave plates on the BS output ports 3 and 4:
\begin{eqnarray}
|\Psi^{1\pm}_{s3,as4}\rangle=\frac{1}{\sqrt{2}}\big(|H,\omega_{s}+\delta\rangle_3|V,\omega_{as}\rangle_4 \nonumber \\
\pm|V,\omega_{s}\rangle_3|H,\omega_{as}+\delta\rangle_4\big),
\label{eq:Hyperentanglement81}
\end{eqnarray}
\begin{eqnarray}
|\Psi^{2\pm}_{s3,as4}\rangle=\frac{1}{\sqrt{2}}\big(|V,\omega_{s}+\delta\rangle_3|H,\omega_{as}\rangle_4 \nonumber \\
\pm|H,\omega_{s}\rangle_3|V,\omega_{as}+\delta\rangle_4\big),
\label{eq:Hyperentanglement82}
\end{eqnarray}
\begin{eqnarray}
|\Phi^{1\pm}_{s3,as4}\rangle=\frac{1}{\sqrt{2}}\big(|H,\omega_{s}+\delta\rangle_3|H,\omega_{as}\rangle_4 \nonumber \\
\pm|V,\omega_{s}\rangle_3|V,\omega_{as}+\delta\rangle_4\big),
\label{eq:Hyperentanglement83}
\end{eqnarray}
\begin{eqnarray}
|\Phi^{2\pm}_{s3,as4}\rangle=\frac{1}{\sqrt{2}}\big(|V,\omega_{s}+\delta\rangle_3|V,\omega_{as}\rangle_4 \nonumber \\
\pm|H,\omega_{s}\rangle_3|H,\omega_{as}+\delta\rangle_4\big),
\label{eq:Hyperentanglement84}
\end{eqnarray}
Exchanging Stokes and anti-Stokes photons, one can obtain other 8 polarization-frequency hyperentangled quantum states, represented by the detection of anti-Stokes at port 3 and Stokes at port 4 ($t_4-t_3<0$).

\begin{figure}
\includegraphics[width=\linewidth]{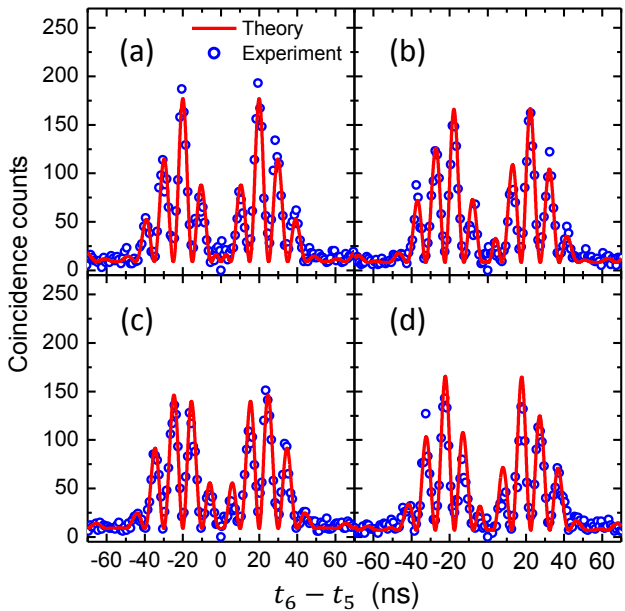}
\caption{\label{fig:FEBeating2} (color online). Two-photon beating for varying phase difference $\theta$ of the frequency entangled state $\frac{1}{\sqrt{2}}\big(|\omega_{s}+\delta,\omega_{as}\rangle+e^{i\theta}|\omega_{s},\omega_{as}+\delta\rangle\big)$: (a) $\theta=0$, (b) $\theta=\pi/2$, (c) $\theta=\pi$, and (d) $\theta=3\pi/2$.}
\end{figure}

Mattle \textit{et al.} show that the polarization birefringent effect is a powerful tool to transfer from one polarization Bell state to others with only linear optics \cite{ZeilingerPRL1996}. In the following, we show that by making use of the polarization-frequency coupling we can continuously vary the relative phase between the two terms in the frequency-entangled Bell state. To achieve this, we project the hyperentangled state in Eq.~(\ref{eq:Hyperentanglement}) to the polarizers P$_3$ and P$_4$ to decouple the polarization degree of freedom from the frequency degree of freedom: $\frac{1}{\sqrt2}(\langle P_3|H\rangle|\langle P_4|V\rangle|\omega_{s}+\delta\rangle_3|\omega_{as}\rangle_4-\langle P_3|V\rangle\langle P_4|H\rangle|\omega_{s}\rangle_3|\omega_{as}+\delta\rangle_4)$ . We insert a quarter-wave plate before the half-wave plate in the BS output port 2 to construct a complex polarizer P$_3$=$\frac{1}{\sqrt{2}}$(H$- e^{-i\theta}$V). P$_4$ remains a linear polarizer P$_4$=$\frac{1}{\sqrt{2}}$(H+V). Then the frequency-entangled state at output ports 5 and 6 [see Fig.~\ref{fig:ExperimetalSetup}(a)] is expressed as
\begin{eqnarray}
\frac{1}{\sqrt2}\big(|\omega_{s}+\delta,\omega_{as}\rangle+e^{i\theta}|\omega_{s},\omega_{as}+\delta\rangle\big).
\label{eq:FE1}
\end{eqnarray}
The Glauber correlation function becomes
\begin{eqnarray}
G^{(2)}_{56}(\tau')=\frac{1}{8}G^{(2)}_0(\tau')\big[1+\cos(\delta \tau'-\theta)\big],
\label{eq:FEG21}
\end{eqnarray}
where $\tau'=t_6-t_5$. The factor 1/8 takes into account the BS and polarizer projection losses. The experimental results (circular data) of phase-shifted two-photon beating at different phase $\theta$ are shown in Fig.~\ref{fig:FEBeating2}, agreeing well with the theory (solid curves). The visibilities of the normalized two-photon beating signal of these measurements are $77\pm5$\%, $78\pm5$\%, $80\pm4$\%,and $77\pm5$\%, for Figs. \ref{fig:FEBeating2}(a), (b), (c), and (d), respectively, which are all beyond the requirement of violating the Bell-CHSH inequality.

In summary, we demonstrate the generation of polarization-frequency coupled hyperentanglement for narrowband biphotons produced from the right-angle SFWM in laser cooled atoms, making use of an AOM frequency shifter and linear optics. As the polarization-frequency degrees of freedom are decoupled, we show that the scheme is robust in creating both polarization Bell states and frequency Bell states, as confirmed by the polarization quantum-state tomography and the two-photon temporal quantum beating. Making use of this polarization-frequency coupling effect followed by a decoupling process, we can transfer the phase difference between the two polarization modes to their entangled frequency modes. We use this technique to produce the frequency entangled Bell state whose phase difference between the two frequency bases can be continuously varied by properly adjusting the wave plates. As compared to the recent demonstration of polarization entanglement for SFWM biphotons \cite{YanPRL2014}, this scheme is much simpler. Moreover, we generate frequency Bell state with an AOM frequency shifter and linear optics. Our demonstration paves the way toward engineering photonic entanglements in polarization and frequency Hilbert spaces.

The work was supported by the Hong Kong Research Grants Council (Project No. 16301214). C. S. acknowledges support from the Undergraduate Research Opportunities Program at the Hong Kong University of Science and Technology.


\end{document}